\documentclass{article}
\usepackage{amsmath,amssymb,cite}

\input{tcilatex}
\begin{document}

\begin{titlepage}
\vspace*{3cm}
\begin{center}
{\Large \textsf{\textbf{Hartmann potential with a minimal length and generalized recurrence relations for matrix elements}}}
\end{center}
\vskip 5mm
\begin{center}
{\large \textsf{Lamine Khodja$^{a,b,}$\footnote{Email: lamine.khodja@yahoo.fr, khodja.lamine@univ-ouagla.dz}, Mohamed Achour$^{b}$ and Slimane Zaim$^{c}$}}\\
\vskip 5mm
$^{a}$D\'{e}partement de Physique, Facult\'{e} des Math\'{e}matiques et des
Sciences de la Mati\`{e}re, Universit\'{e} Kasdi Merbah -- Ouargla, Algeria.\\
$^{b}$Laboratoire de Physique Th\'{e}orique, Facult\'{e} des Sciences Exactes, Universit\'{e} de Bejaia, 06000 Bejaia, Algeria.\\
$^{c}$D\'{e}partement de Physique, Facult\'{e} des Sciences  de la Mati\`{e}re,\\
Universit\'{e} de Batna--1 Hadj Lakhdar, Algeria.\\
\end{center}
\vskip 2mm
\begin{center}{\large\textsf{\textbf{Abstract}}}\end{center}
\begin{quote}
In this work we study the Schr\"{o}dinger equation in the presence of the Hartmann potential with a generalized uncertainty principle. We pertubatively obtain the matrix elements of the hamiltonian at first order in the parameter of deformation $\beta$ and show that some degenerate states are removed. We give analytic expressions for the solutions of the diagonal matrix elements. Finally, we derive a generalized recurrence formula for the angular average values.
\end{quote}
\vspace*{2cm}
\noindent\textbf{\sc Pacs numbers}: 02.30.Gp, 31.15.-p, 31.15.xp
\end{titlepage}

\section{Introduction}

In recent years many arguments have been suggested to motivate a modified
Heisenberg algebra in quantum mechanics such as theories of quantum gravity
and string theory, which lead to the existence of a minimal observable
length expected to be of order of the Planck length \cite%
{Stetko1,Stetko2,Stetko3,Haouat5,Haouat6,Haouat7,Haouat8,Haouat9,Haouat10,Haouat11,Haouat12}%
. The minimal length can be obtained from the deformed canonical commutation
relation between position and momentum operators \cite{Haouat00, Stetko4,
Stetko5, Stetko6, Stetko7, Stetko8}: 
\begin{equation}
\left[ X,P\right] =i\hbar \left( 1+\beta P^{2}\right) ,  \label{a}
\end{equation}%
where $\beta $ is a positive parameter of deformation. This commutation
relation implies the following generalized uncertainty principle (GUP): 
\begin{equation}
\Delta X\Delta P\geq \frac{\hbar }{2}\left( 1+\beta \left( \Delta P\right)
^{2}\right) ,
\end{equation}%
which corresponds to a minimal length $\left( \Delta X\right) _{\min }=\hbar 
\sqrt{\beta }$.

There has recently been a lot of interest in the study of quantum mechanics
problems in the presence of a minimal length with various potentials such as
the harmonic oscillator, the Coulomb and Yukawa potentials, the Woods-Saxon
potential, the Kratzer potential, $\hdots$ \cite{merad, chargui, Haouat18,
Haouat19, Haouat20, Haouat21, Haouat22, Haouat23, Haouat24, Haouat25,
Haouat29}. The first author to use the standard perturbation theory to solve
the Schr\"{o}dinger equation of the central potentials in the presence of a
minimal length was Brau in Ref. \cite{Haouat19}, where energy-level
corrections were calculated and splitting of degenerate levels were found to
occur. On the other hand the solutions of the Schr\"{o}dinger equation with
Hartmann potential $V\left( r,\theta \right) =\eta \sigma ^{2}\left( \frac{%
e^{2}}{r}+\frac{q\hbar ^{2}}{2\mu r^{2}\sin ^{2}\theta }\right) $, where $%
\eta $ and $\sigma $ are positive real numbers with values ranging from
about 1 to 10, and $q$ is a real parameter, is well known \cite{Chang0,
Chang00, Chang4, Chang5, Chang6, Chang7, Chang8}. This potential has been
introduced in \cite{Chang1, Chang2, Chang3} to describe ring-shaped
molecules. When $q=0$ and $\eta \sigma ^{2}=Z$ the Hartmann potential
reduces to the Coulomb potential.

The purpose of this paper is to study the extension of the Schr\"{o}dinger
equation with the Hartmann potential in the presence of a minimal length. In
section 2 we use the first-order perturbation theory to give the general
form of the hamiltonian matrix elements and, for a particular case, we show
that the degeneracy is completely removed. In section 3 we give an explicit
analytical expression of the diagonal matrix elements and show that the
splitting of the degenerate energy levels also occurs. In section 4 we
provide a general recurrence formula for the angular part. Finally, in the
last section, we draw our conclusion.

\section{Hamiltonian matrix elements}

To calculate the hamiltonian matrix elements for a Hartmann potential in the
presence of a minimal length, we solve the corresponding Schr\"{o}dinger
equation: 
\begin{equation}
\left[\frac{\hat{P}^2}{2\mu}+V\left(\hat{r},\hat{\theta}\right)\right]%
\psi\left(\vec{r}\right) =E^{\left( \beta\right) }\psi\left( \vec{r}\right).
\label{ref:a}
\end{equation}
We choose to work with the following representation that verifies the
relation \eqref{a} to first order in $\beta$: 
\begin{align}
\hat{X}_{i}\psi \left( \vec{r}\right) & =x_{i}\psi \left( \vec{r}\right), \\
\hat{P}_{i}\psi \left( \vec{r}\right) & =p_{i}\left( 1+\beta p^{2}\right)
\psi \left( \vec{r}\right),\qquad p_{i}=i\hbar \frac{\partial }{\partial
x^{i}}\,.
\end{align}

To first order in $\beta$ the Schr\"{o}dinger equation \eqref{ref:a} can be
written as: 
\begin{equation}
\left[\frac{p^{2}}{2\mu}+\frac{\beta p^{4}}{\mu}+V\left( r,\theta\right)%
\right] \hat{\psi}\left( r,\theta,\varphi\right) =E^{\left( \beta\right) }%
\hat{\psi}\left( r,\theta,\varphi\right).  \label{a1}
\end{equation}
In this equation the Hartmann potential in the presence of a minimal length
appears within the perturbation term: 
\begin{equation*}
\frac{\beta p^{4}}{\mu}
\end{equation*}

To investigate the correlations we use the first-order perturbation theory.
For $\beta =0$ the spectrum of equation \eqref{a1} and the corresponding
wave functions are well-known and are given by \cite{Chang0,yasuk,Sameer}: 
\begin{equation}
\psi\left(r,\theta,\varphi\right)=\frac{1}{r}R\left(r\right)\Theta\left(%
\theta\right)\Phi\left(\phi\right),
\end{equation}
where 
\begin{align}
&\Phi\left(\phi\right)=\frac{1}{\sqrt{2\pi}}\exp\left(im\phi\right),\qquad
m=0,\pm1,\pm2,\pm3,\hdots  \notag \\
&\Theta_{nm}\left( x\right)=\frac{\Gamma\left(2k+1\right)}{%
\Gamma\left(k+1\right)}\sqrt{\frac{\left( 2n+2k+1\right)}{%
2^{2k+1}\Gamma\left(n+2k+1\right)}}\left(1-x^{2}\right)^{k/2}C_{n}^{%
\left(k+1/2\right)} \left(x\right),\,x=\cos\theta  \notag \\
&R_{Nnm}\left(r\right)=\left(\frac{\mu\eta\sigma^{2}e^{2}}{\hbar^2n^\prime}%
\right)^{1/2}\left[\frac{\left(n^{\prime}-l-1\right)!}{n^\prime\Gamma%
\left(n^{\prime}+l+1\right)}\right]^{1/2}\left(\frac{ 2\mu\eta\sigma^{2}e^{2}%
}{\hbar^{2}n^{\prime}}r\right) ^{l+1}  \notag \\
&\qquad\qquad\times \exp\left(-\frac{\mu\eta\sigma^{2}e^{2}}{%
\hbar^{2}n^{\prime}}r\right) L_{N}^{\left(2l+1\right) }\left( \frac{%
2\mu\eta\sigma^{2}e^{2}}{\hbar^{2}n^{\prime}}r\right),  \notag \\
&k =\sqrt{m^{2}+\frac{q\eta\hbar^{2}\sigma^{2}}{2\mu}}\,,\quad
n^{\prime}=N+l+1\,,\quad l=n+k\,,\quad N,n=0,1,2,3,\hdots
\end{align}
with $N$ being the radial quantum number, $L_{n}^{\left( \nu \right)
}\left(x\right)$ and $C_{n}^{\left(\nu\right)}\left(x\right)$ respectively
stand for the associated Laguerre and the Gegenbauer (ultraspherical)
polynomials. The orthogonality conditions for these functions are: 
\begin{align}
&\int_{-1}^{1}dx\left(1-x^{2}\right)^{\nu-1/2}\left[C_{n}^{\left(\nu\right)}%
\left(x\right)\right]^2=\frac{\pi2^{1-2\nu }\Gamma\left(n+2\nu\right)}{%
n!\left(n+\nu\right)\left[\Gamma\left(\nu\right)\right]^2}\,,  \notag \\
&\int_{-1}^{1}dxe^{-x}x^{\nu }\left[L_{n}^{\left(\nu\right)}\left(x\right)%
\right]^2=\frac{\Gamma\left(\nu+n+1\right)}{n!}\,.
\end{align}%
The energy eigenvalues are: 
\begin{equation}
E_{Nnm}^{\left( 0\right) }=-\frac{\mu \left( \eta \sigma ^{2}\right)^2e^{4}}{%
2\hbar ^{2}}\left[ N+n+\sqrt{m^{2}+\frac{q\eta \sigma ^{2}\hbar^{2}}{2\mu }}%
+1\right] ^{-2}.
\end{equation}

The first-order perturbation theory gives the matrix element of the
hamiltonian operator \eqref{a1} up to first order in $\beta $ as follows 
\cite{Haouat19}: 
\begin{align}
&\frac{\beta }{\mu }\left\langle N_{1}n_{1}m_{1}\right\vert p^{4}\left\vert
N_{2}n_{2}m_{2}\right\rangle=4\mu \beta \left[ \left(
E_{N_{12}n_{2}m_{2}}^{\left( 0\right) }\right) ^{2}\delta
_{N_{1}N_{2}}\delta _{n_{1}n_{2}}\delta_{m_{1}m_{2}} \right.  \notag \\
&\left.-2E_{N_{2}n_{2}m_{2}}^{\left( 0\right) }\left\langle
N_{1}n_{1}m_{1}\right\vert V\left( r,\theta \right) \left\vert
N_{2}n_{2}m_{2}\right\rangle+ \left\langle N_{1}n_{1}m_{1}\right\vert \left(
V\left( r,\theta \right) \right) ^{2}\left\vert N_{2}n_{2}m_{2}\right\rangle 
\right] ,  \label{aa}
\end{align}
where 
\begin{equation}
V\left( r,\theta\right) =\eta\sigma^{2}\left( \frac{e^{2}}{r}+\frac {%
q\hbar^{2}}{2\mu r^{2}\sin^{2}\theta}\right).
\end{equation}
Each of these terms can be written as 
\begin{align}
\left\langle N_{1}n_{1}m_{1}\right\vert \frac{1}{r^{s}\sin ^{2t}\theta}%
\left\vert N_{2}n_{2}m_{2}\right\rangle &=\int_{0}^{\infty
}R_{N_{1}n_{1}m_{1}}\left( r\right) R_{N_{2}n_{2}m_{2}}\left(
r\right)r^{-s}dr  \notag \\
&\quad \times \int_{-1}^{1}dx\left( 1-x^{2}\right) ^{-t}\Theta
_{n_{1}m_{1}}\left(x\right) \Theta _{n_{2}m_{2}}\left( x\right)
\end{align}
For the radial part, the first integral has been evaluated in \cite{Chang00}
and its expression is given by: 
\begin{align}
&\left\langle N_{1}n_{1}m_{1}\right\vert r^{s}\left\vert
N_{2}n_{2}m_{2}\right\rangle=\int_{0}^{\infty
}R_{N_{1}n_{1}m_{1}}\left(r\right) R_{N_{2}n_{2}m_{2}}\left( r\right) r^{s}dr
\notag \\
&=\frac{\eta \sigma ^{2}}{n_{1}n_{2}}\sqrt{\frac{N_{1}!N_{2}!}{\Gamma\left(
2l_{1}+N+2\right) \Gamma \left( 2l_{2}+N+2\right) }}\left( \frac{2\eta
\sigma ^{2}}{n_{1}^{\prime }}\right) ^{l_{1}+1}\left( \frac{2\eta \sigma ^{2}%
}{n_{2}^{\prime }}\right) ^{l_{2}+1}  \notag \\
&\times\left[ \eta \sigma ^{2}\left( 1/n_{1}^{\prime }+1/n_{2}^{\prime
}\right)\right] ^{-s-l_{1}-l_{2}-3} \sum_{m_{1}}^{N}\sum_{m_{2}}^{N}\frac{%
\left( -1\right) ^{2N+m_{2}}}{ m_{1}!m_{2}!}\left( \frac{1/n_{1}^{\prime
}-1/n_{2}^{\prime }}{1/n_{1}^{\prime }+1/n_{2}^{\prime }}\right)
^{m_{1}+m_{2}}  \notag \\
& \times \Gamma \left(
l_{1}+l_{2}+s+m_{1}+m_{2}+3\right)\sum_{m_{3}=0}^{t}\left( 
\begin{array}{c}
l_{1}+l_{2}+s+m_{2}+1 \\ 
N-m_{1}-m_{3}%
\end{array}%
\right)  \notag \\
&\times \left(%
\begin{array}{c}
l_{1}+l_{2}+s+m_{1}+1 \\ 
N-m_{2}-m_{3}%
\end{array}
\right) \left( 
\begin{array}{c}
l_{1}+l_{2}+s+m_{1}+m_{2}+m_{3}+2 \\ 
m_{3}%
\end{array}
\right) ,  \label{ab}
\end{align}
with $t=\min\left(N-m_{1},N-m_{2}\right) $ and $s<l_{1}+l_{2}+3$. The only
values of $s$ which contribute in the calculation of the matrix elements
given in the expression \eqref{aa} are $s=\left\{ -1,-2,-3,-4\right\} $.

For the angular part we use the following integral of the product of two
Gegenbauer polynomials \cite{Wassouli} with $\theta >-1/2$, $\mu >-1/2$, and 
$\lambda >-1/2$: 
\begin{align}
&\int_{-1}^{1}C_{l}^{\theta }\left( x\right) C_{m}^{\mu }\left(
x\right)\left( 1-x^{2}\right) ^{\lambda -1/2}dx=\frac{\pi 2^{1-2\lambda }}{%
\Gamma\left( \mu \right) \Gamma \left( \theta \right) }\sum_{k=0}^{[l/2]}%
\left[ \frac{\left( l-2k+\lambda \right) }{\left( l-2k\right) !k!s!}\right. 
\notag \\
&\times\left.\frac{\Gamma \left( l-\theta -k\right) }{\Gamma \left(
l+\lambda-k+1\right) }\frac{\Gamma \left( l-2k+2\lambda \right) \Gamma
\left( m+\mu-s\right) }{\Gamma \left( m+\lambda -s+1\right)} \left( \theta
-\lambda \right) _{k}\left( \mu -\lambda\right) _{s}\right] ,  \label{cc}
\end{align}
where $m=l-2k+2s$, with $[m/2]\geq s\in N$, and $\left(z\right)_{n}$ is the
Pochhammer symbol \cite{Abramowitz}: 
\begin{equation}
\left(z\right) _{n}=z\left( z+1\right) \hdots \left( z+n-1\right) =\frac{%
\Gamma\left( z+n\right) }{\Gamma\left( z\right)}.
\end{equation}
The integral \eqref{cc} vanishes for odd values of $l+m$. We thus have the
following special integrals: 
\begin{align}
&\int_{-1}^{1}dx\left( 1-x^{2}\right) ^{-1}\Theta
_{n_{1}m_{1}}\left(x\right) \Theta _{n_{2}m_{2}}\left( x\right) =\left\vert
\cos \frac{\pi }{2}\left( n_{1}+n_{2}\right) \right\vert  \notag \\
&\times \frac{\Gamma \left( 2k_{1}+1\right)}{\Gamma \left( k_{1}+1\right) }%
\frac{\Gamma \left( 2k_{1}+1\right)}{\Gamma\left( k_{1}+1\right)}\left[\frac{%
\left( 2n_{1}+2k_{1}+1\right)}{2^{2\left(k_{1}+k_{2}+1\right)} \Gamma\left(
n_{1}+2k_{1}+1\right) }\frac{\left(2n_{2}+2k_{2}+1\right) }{\Gamma \left(
n_{2}+2k_{2}+1\right) }\right] ^{1/2}  \notag \\
&\times\frac{\pi 2^{2-k_{1}-k_{2}}}{\Gamma\left(k_{1}+1/2\right)\Gamma%
\left(k_{2}+1/2\right) }\sum_{p=0}^{[n_{1}/2]}\left[\frac{\left[
n_{1}-2p+\left(k_{1}+k_{2}-1\right)/2\right]}{\left(n_{1}-2p\right)!p!s!}%
\right.  \notag \\
&\times\frac{\Gamma \left( n_{1}-2p+k_{1}+k_{2}-1\right) \Gamma
\left(n_{2}+k_{1}-s+1/2\right) }{\Gamma \left( n_{2}+\frac{k_{1}+k_{2}+1}{2}%
-s\right) }  \notag \\
&\left.\times \frac{\Gamma \left( n_{1}-k_{1}-p-1/2\right) }{\Gamma
\left(n_{1}+\frac{k_{1}+k_{2}+1}{2}-p\right) }\left( \frac{k_{1}-k_{2}}{2}%
+1\right) _{p}\left(\frac{k_{2}-k_{1}}{2}+1\right)_{s}\right].
\end{align}
and 
\begin{align}
&\int_{-1}^{1}dx\left(
1-x^{2}\right)^{-2}\Theta_{n_{1}m_{1}}\left(x\right)\Theta_{n_{2}m_{2}}%
\left(x\right)=\left\vert\cos\frac{\pi }{2}\left(n_{1}+n_{2}\right)\right%
\vert\times  \notag \\
&\times\frac{\Gamma \left( 2k_{1}+1\right)}{\Gamma \left( k_{1}+1\right) }%
\frac{\Gamma \left( 2k_{1}+1\right) }{\Gamma\left( k_{1}+1\right)}\left[ 
\frac{\left( 2n_{1}+2k_{1}+1\right) }{2^{2\left(k_{1}+k_{2}+1\right) }\Gamma
\left( n_{1}+2k_{1}+1\right) }\frac{\left(2n_{2}+2k_{2}+1\right) }{\Gamma
\left( n_{2}+2k_{2}+1\right) }\right] ^{1/2}  \notag \\
&\times\frac{\pi 2^{4-k_{1}-k_{2}}}{\Gamma \left( k_{1}+1/2\right) \Gamma
\left(k_{2}+1/2\right) }\sum_{p=0}^{[n_{1}/2]}\left[ \frac{\left[
n_{1}-2p+\left(k_{1}+k_{2}-3\right) /2\right] }{\left( n_{1}-2p\right) !p!s!}%
\right.  \notag \\
&\times \frac{\Gamma \left( n_{1}-2p+k_{1}+k_{2}-3\right) \Gamma
\left(n_{2}+k_{1}-s+1/2\right) }{\Gamma \left( n_{2}+\frac{k_{1}+k_{2}-1}{2}%
-s\right) }  \notag \\
&\times \left.\frac{\Gamma \left( n_{1}-k_{1}-p-1/2\right) }{\Gamma
\left(n_{1}+\frac{k_{1}+k_{2}-1}{2}-p\right) }\left( \frac{k_{1}-k_{2}}{2}%
+2\right) _{p}\left( \frac{k_{2}-k_{1}}{2}+2\right) _{s}\right].
\end{align}

Finally, the general form of the hamiltonian matrix elements in \eqref{aa}
is given by the expression: 
\begin{align}
&\frac{\beta }{4\mu }\left\langle N_{1}n_{1}m_{1}\right\vert p^{4}\left\vert
N_{2}n_{2}m_{2}\right\rangle =\frac{\mu ^{2}\left( \eta \sigma ^{2}\right)
^{4}e^{8}}{4\hbar ^{4}}\left( n_{2}^{\prime }\right) ^{-4}\delta
_{N_{1}N_{2}}\delta _{n_{1}n_{2}}\delta _{m_{1}m_{2}}  \notag \\
&+\frac{\mu \left( \eta \sigma ^{2}\right) ^{4}e^{6}}{\hbar ^{2}}\left(
n_{2}^{\prime }\right) ^{-2}\left\langle N_{1}n_{1}m_{1}\right\vert
r^{-1}\left\vert N_{2}n_{2}m_{2}\right\rangle  \notag \\
& +\left( \eta \sigma ^{2}\right) ^{2}e^{4}\left\langle
Nn_{1}m_{1}\right\vert r^{-2}\left\vert N_{2}n_{2}m_{2}\right\rangle  \notag
\\
&+\left[\frac{\left( \eta \sigma ^{2}\right) ^{4}e^{6}q}{2}\left(
n_{2}^{\prime }\right) ^{-2}\left\langle N_{1}n_{1}m_{1}\right\vert
r^{-2}\left\vert N_{2}n_{2}m_{2}\right\rangle\right.  \notag \\
&\qquad+\left.\frac{\left( \eta \sigma ^{2}\right) e^{2}q\hbar ^{2}}{\mu }%
\left\langle N_{1}n_{1}m_{1}\right\vert r^{-3}\left\vert
N_{2}n_{2}m_{2}\right\rangle \right]  \notag \\
&\times \frac{\Gamma \left( 2k_{1}+1\right) }{\Gamma \left( k_{1}+1\right) }%
\frac{ \Gamma \left( 2k_{1}+1\right) }{\Gamma \left( k_{1}+1\right) }\left[ 
\frac{\left( 2n_{1}+2k_{1}+1\right) }{2^{2\left( k_{1}+k_{2}+1\right)
}\Gamma \left( n_{1}+2k_{1}+1\right) }\frac{\left( 2n_{2}+2k_{2}+1\right) }{%
\Gamma \left( n_{2}+2k_{2}+1\right) }\right] ^{1/2}  \notag \\
& \times\frac{\pi 2^{2-k_{1}-k_{2}}}{\Gamma \left( k_{1}+1/2\right) \Gamma
\left( k_{2}+1/2\right) }\left\vert \cos \frac{\pi }{2}\left(
n_{1}+n_{2}\right) \right\vert  \notag \\
&\times \sum_{p=0}^{[n_{1}/2]}\left[ \frac{\left[ n_{1}-2p+\left(
k_{1}+k_{2}-1\right) /2\right] }{\left( n_{1}-2p\right) !p!s!}\right.  \notag
\\
& \times \frac{\Gamma \left( n_{1}-2p+k_{1}+k_{2}-1\right) \Gamma \left(
n_{2}+k_{1}-s+1/2\right) }{\Gamma \left( n_{2}+\frac{k_{1}+k_{2}+1}{2}
-s\right) }  \notag \\
&\times \left. \frac{\Gamma \left( n_{1}-k_{1}-p-1/2\right) }{\Gamma \left(
n_{1}+\frac{k_{1}+k_{2}+1}{2}-p\right) }\left( \frac{k_{1}-k_{2}}{2}%
+1\right) _{p}\left( \frac{k_{2}-k_{1}}{2}+1\right) _{s}\right]  \notag \\
& +\left\vert \cos \frac{\pi }{2}\left( n_{1}+n_{2}\right) \right\vert
\left( \eta \sigma ^{2}\right) ^{2}\frac{q^{2}\hbar ^{4}}{4\mu ^{2}}\frac{%
\Gamma \left( 2k_{1}+1\right) }{\Gamma \left( k_{1}+1\right) }\frac{\Gamma
\left( 2k_{1}+1\right) }{\Gamma \left( k_{1}+1\right) }  \notag \\
&\times \left[ \frac{\left( 2n_{1}+2k_{1}+1\right) }{2^{2\left(
k_{1}+k_{2}+1\right) }\Gamma \left( n_{1}+2k_{1}+1\right) }\frac{\left(
2n_{2}+2k_{2}+1\right) }{\Gamma \left( n_{2}+2k_{2}+1\right) }\right] ^{1/2}
\notag \\
&\times \frac{\pi 2^{4-k_{1}-k_{2}}}{\Gamma \left( k_{1}+1/2\right) \Gamma
\left(k_{2}+1/2\right) }\sum_{p=0}^{[n_{1}/2]}\left[ \frac{\left[
n_{1}-2p+\left(k_{1}+k_{2}-3\right) /2\right] }{\left( n_{1}-2p\right) !p!s!}%
\right.  \notag \\
& \times \frac{\Gamma \left( n_{1}-2p+k_{1}+k_{2}-3\right) \Gamma
\left(n_{2}+k_{1}-s+1/2\right) }{\Gamma \left( n_{2}+\frac{k_{1}+k_{2}-1}{2}%
-s\right) }  \notag \\
& \left. \times \frac{\Gamma \left( n_{1}-k_{1}-p-1/2\right) }{\Gamma \left(
n_{1}+\frac{k_{1}+k_{2}-1}{2}-p\right) }\left( \frac{k_{1}-k_{2}}{2}%
+2\right) _{p}\left( \frac{k_{2}-k_{1}}{2}+2\right) _{s}\right]  \notag \\
& \times\left\langle N_{1}n_{1}m_{1}\right\vert r^{-4}\left\vert
N_{2}n_{2}m_{2}\right\rangle\,,
\end{align}
where $\left\langle N_{1}n_{1}m_{1}\right\vert r^{s}\left\vert
N_{2}n_{2}m_{2}\right\rangle $ are given by replacing $s=\left\{-1,-2,-3,-4%
\right\} $ in \eqref{ab}.

\subsection*{Example: The states $\boldsymbol{\left\vert 010\right\rangle }$
and $\boldsymbol{\left\vert 100\right\rangle }$}

In the ordinary case (i.e. $\beta =0$), $\left\vert 010\right\rangle $ and $%
\left\vert 100\right\rangle $ are two degenerate states. It is clear that
the matrix \eqref{aa} is actually diagonal ($\left\langle 100\right\vert
p^{4}\left\vert 010\right\rangle =0$) which can be seen by using the
expressions: 
\begin{eqnarray}
L_{0}^{\left( \alpha \right) }\left( x\right) &=&1\ \ \ \ \ ,\ \ \ \ \
\,L_{1}^{\left( \alpha \right) }\left( x\right) =-x+\alpha +1  \label{bb1} \\
C_{0}^{\left( \alpha \right) }\left( x\right) &=&1\ \ \ \ \ ,\ \ \ \ \
C_{1}^{\left( \alpha \right) }\left( x\right) =2\alpha x\,  \label{bb2}
\end{eqnarray}%
From equation \eqref{aa}, a straightforward calculation gives the energy
corrections at first order in the parameter $\beta $ as follows: 
\begin{align}
& \Delta E_{010}=4\mu ^{2}\left[ \left( E_{010}^{\left( 0\right) }\right)
^{2}-\frac{abAE_{010}^{\left( 0\right) }\left( 2k_{0}+1\right) ^{2}}{k_{0}+2}%
\left( \frac{2^{2k_{0}+1}\left[ \Gamma \left( k_{0}+1\right) \right] ^{2}}{%
\Gamma \left( 2k_{0}+2\right) }-\frac{2^{2k+3}\left[ \Gamma \left(
k_{0}+2\right) \right] ^{2}}{\Gamma \left( 2k_{0}+4\right) }\right) \right. 
\notag \\
& -\frac{a^{2}bBE_{010}^{\left( 0\right) }\left( 2k_{0}+1\right) ^{2}}{%
\left( k_{0}+2\right) \left( 2k_{0}+3\right) }\left( \frac{2^{2k_{0}-1}\left[
\Gamma \left( k_{0}\right) \right] ^{2}}{\Gamma \left( 2k\right) }-\frac{%
2^{2k_{0}+1}\left[ \Gamma \left( k_{0}+1\right) \right] ^{2}}{\Gamma \left(
2k_{0}+2\right) }\right)  \notag \\
& +\frac{a^{2}bA^{2}\left( 2k_{0}+1\right) ^{2}}{2\left( k_{0}+2\right)
\left( 2k_{0}+3\right) }\left( \frac{2^{2k_{0}+1}\left[ \Gamma \left(
k_{0}+1\right) \right] ^{2}}{\Gamma \left( 2k_{0}+2\right) }-\frac{%
2^{2k_{0}+3}\left[ \Gamma \left( k_{0}+2\right) \right] ^{2}}{\Gamma \left(
2k_{0}+4\right) }\right)  \notag \\
& +\frac{a^{4}bB^{2}\left( 2k_{0}+1\right) }{2\left( k_{0}+2\right) \left(
2k_{0}+3\right) \left( 2k_{0}+2\right) }\left( \frac{2^{2k_{0}-1}\left[
\Gamma \left( k_{0}-1\right) \right] ^{2}}{\Gamma \left( 2k_{0}-2\right) }-%
\frac{2^{2k_{0}+1}\left[ \Gamma \left( k_{0}\right) \right] ^{2}}{\Gamma
\left( 2k_{0}\right) }\right)  \notag \\
& \left. +\frac{a^{3}bAB\left( 2k_{0}+1\right) ^{2}}{\left( k_{0}+2\right)
\left( 2k_{0}+3\right) \left( 2k_{0}+2\right) }\left( \frac{2^{2k_{0}-1}%
\left[ \Gamma \left( k_{0}\right) \right] ^{2}}{\Gamma \left( 2k_{0}\right) }%
-\frac{2^{2k_{0}+1}\left[ \Gamma \left( k_{0}+1\right) \right] ^{2}}{\Gamma
\left( 2k_{0}+2\right) }\right) \right]  \notag \\
& \Delta E_{100}=4\mu ^{2}\left[ \left( E_{100}^{\left( 0\right) }\right)
^{2}-\frac{abAE_{100}^{\left( 0\right) }}{\left( 2+k_{0}\right) }\frac{%
2^{2k_{0}+1}\left[ \Gamma \left( k_{0}+1\right) \right] ^{2}}{\Gamma \left(
2k_{0}+2\right) }\right.  \notag \\
& -\frac{a^{2}bBE_{100}^{\left( 0\right) }2^{2k_{0}-1}\left[ \Gamma \left(
k_{0}\right) \right] ^{2}}{\left( 2+k_{0}\right) \left( 2k_{0}+1\right)
\Gamma \left( 2k_{0}\right) }+\frac{a^{2}bA^{2}2^{2k_{0}}\left[ \Gamma
\left( k_{0}+1\right) \right] ^{2}}{\left( 2+k_{0}\right) \left(
2k_{0}+1\right) \Gamma \left( 2k_{0}+2\right) }  \notag \\
& \left. +\frac{a^{4}bB^{2}2^{2k_{0}-3}\left( 2k_{0}-2\right) \left(
4+k_{0}\right) \left[ \Gamma \left( k_{0}-1\right) \right] ^{2}}{\left(
2+k_{0}\right) \left( 2k_{0}+3\right) }+\frac{a^{3}bAB2^{2k_{0}}\left[
\Gamma \left( k_{0}\right) \right] ^{2}}{\Gamma \left( 2k_{0}+3\right) }%
\right]
\end{align}%
where 
\begin{align}
E_{100}^{\left( 0\right) }& =E_{010}^{\left( 0\right) }=-\frac{\mu \left(
\eta \sigma ^{2}\right) ^{2}e^{4}}{2\hbar ^{2}}\left[ 2+\sqrt{\frac{q\eta
\sigma ^{2}\hbar ^{2}}{2\mu }}\right] ^{-2},\,A=\eta \sigma ^{2}e^{2},\,B=%
\frac{\eta q\sigma ^{2}\hbar ^{2}}{2\mu }, \\
a& =\frac{2\mu \eta \sigma ^{2}e^{2}}{\hbar ^{2}n^{\prime }}\text{ },\text{ }%
b=\left[ \frac{\Gamma \left( 2k+1\right) }{\Gamma \left( k+1\right) }\right]
^{2}\frac{\left( 2n+2k+1\right) }{2^{2k+1}\Gamma \left( n+2k+1\right) }\text{
\ },\text{ }k_{0}=\sqrt{\frac{q\eta \hbar ^{2}\sigma ^{2}}{2\mu }}.
\end{align}%
Thus, at first order in the parameter $\beta $, the degeneracy of the two
levels $\left\vert 010\right\rangle $ and $\left\vert 100\right\rangle $ is
completely lifted.

\section{Matrix elements for $\boldsymbol{\Delta N=0}$, $\boldsymbol{\Delta
n=0}$ \newline
and $\boldsymbol{\Delta m=0}$}

Taking $N_{1}=N_{2}=N$, $n_{1}=n_{2}=n$ and $m_{1}=m_{2}=m$, we derive the
explicit form of the energy corrections given in equation \eqref{aa}. Using
the relation between the confluent hypergeometric function $F\left(
-n;l+1;x\right) $ and the associated Laguerre polynomials $L_{n}^{\left(
l\right) }(x)$, namely: 
\begin{equation}
L_{n}^{\left( l\right) }(z)=\frac{\Gamma \left( n+l+1\right) }{\Gamma
\left(n+1\right) \Gamma \left( l+1\right) }F(-n;l+1;z)\,,
\end{equation}
where $z=ar$ and $a=\frac{2\mu \eta \sigma ^{2}e^{2}}{\hbar ^{2}n^{\prime }}$%
, and using the integral: 
\begin{align}
&\int_{0}^{\infty}z^{l-1}e^{-z}\left[ F(-n;\gamma;z)\right] ^{2}dx =\frac{%
n!\Gamma(l)}{\gamma\left( \gamma+1\right) \cdots\left( \gamma+n-1\right) }%
\bigg\{ 1+  \notag \\
&\left.\frac{n\left( \gamma-l-1\right) \left( \gamma-l\right) }{1^{2}\gamma}+%
\frac{n\left( n-1\right) \left( \gamma-l-2\right) \left(\gamma-l-1\right)
\left( \gamma-l\right) \left( \gamma-l+1\right) }{1^{2}2^{2}\gamma\left(
\gamma+1\right) }+\cdots\right.  \notag \\
& \left. \cdots+\frac{n\left( n-1\right) \cdots1\left(
\gamma-l-n\right)\cdots\left( \gamma-l+n-1\right) }{1^{2}2^{2}\cdots
n^{2}\gamma\left(\gamma+1\right) \cdots\left( \gamma+n-1\right) }\right\},
\end{align}
we obtain the following average values for the radial part: 
\begin{align}
&\langle Nnm \mid r^{-1}\mid Nnm\rangle=\int_{0}^{\infty}\left[
R\left(r\right) \right] ^{2}r^{-1}dr=\frac{a}{2n^{\prime}}\,, \\
&\langle Nnm \mid r^{-2}\mid Nnm\rangle=\int_{0}^{\infty}\left[
R\left(r\right) \right] ^{2}r^{-2}dr=\frac{1}{2l+1}\frac{a^{2}}{2n^{\prime}}%
\,, \\
&\langle Nnm \mid r^{-3}\mid Nnm\rangle=\int_{0}^{\infty}\left[
R\left(r\right) \right] ^{2}r^{-3}dr=\frac{a^{3}}{2l\left( 2l+1\right)
\left( 2l+2\right) }=\frac {a^{3}\Gamma\left( 2l\right) }{\Gamma\left(
2l+3\right)}\,,
\end{align}
\begin{align}
&\langle Nnm \mid r^{-4}\mid Nnm^{\prime}\rangle=\int_{0}^{\infty}\left[%
R\left( r\right) \right] ^{2}r^{-4}dr  \notag \\
&=\frac{a^{4}}{n^{\prime}}\left( \frac{3\left( n^{\prime}\right)^{2}-l\left(
l+1\right) }{\left( 2l-1\right) 2l\left( 2l+1\right) \left(2l+2\right)
\left( 2l+3\right) }\right)  \notag \\
&=\frac{a^{4}\left[ 3\left( n^{\prime}\right) ^{2}-l\left( l+1\right) \right]
}{n^{\prime}}\frac{\Gamma\left( 2l-1\right) }{\Gamma\left(2l+4\right) }
\end{align}

Now, we evaluate the following integral of the angular functions: 
\begin{align}
&\int_{-1}^{1}dx\left( 1-x^{2}\right) ^{-t}\left[ \Theta\left( x\right)%
\right] ^{2}=\frac{\left( 2n+2k+1\right) }{2^{2k+1}\Gamma\left(n+2k+1\right) 
}  \notag \\
&\times\left[ \frac{\Gamma\left( 2k+1\right) }{\Gamma\left( k+1\right) }%
\right] ^{2}\int_{-1}^{1}dx\left( 1-x^{2}\right) ^{k-t}\left[C_{n}^{\left(
k+1/2\right) }\left( x\right) \right] ^{2}
\end{align}
We use the following expression given in Ref. \cite{Brychkov1}: 
\begin{align}
&\int_{0}^{\pi }d\theta \sin ^{\nu }\theta \sin \left( \gamma \theta \right) 
\left[ C_{n}^{\left( \lambda \right) }\left( \sqrt{1+\rho \sin ^{2}\theta }%
\right) \right] ^{2} =\frac{2^{-\nu }\pi \Gamma \left( \nu +1\right)
\left(2\lambda \right) _{n}^{2}}{\left( n!\right) ^{2}\Gamma \left( \frac{%
\nu-\gamma }{2}+1\right) \Gamma \left( \frac{\nu +\gamma }{2}+1\right)} 
\notag \\
&\times\sin \left( \frac{\gamma \pi }{2}\right)\, _5F_4\left(%
\begin{array}{c}
-n,\lambda ,2\lambda +n,\frac{\nu +1}{2},1+\frac{\nu }{2};-\rho \\ 
\lambda +\frac{1}{2},2\lambda ,\frac{\nu -\gamma }{2}+1,\frac{\nu +\gamma }{2%
}+1%
\end{array}%
\right),\qquad \left( \mathrm{Re}\nu >-1\right)
\end{align}
with $_pF_q\binom{a_{1},a_{2},\hdots a_{p};x}{b_{1},b_{2},\hdots,b_{q}}$
being the hypergeometric function defined as: 
\begin{align}
_pF_q\binom{a_{1},a_{2},\hdots,a_{p};x}{b_{1},b_{2},\hdots,b_{q}} &
=\,_pF_q\left( a_{1},a_{2},\hdots,a_{p};b_{1},b_{2},\hdots,b_{q};x\right) 
\notag \\
&=\sum\limits_{k=0}^{\infty}\frac{\left( a_{1}\right) _{k}\left(a_{2}\right)
_{k}\hdots\left( a_{p}\right) _{k}}{\left( b_{1}\right)
_{k}\left(b_{2}\right) _{k}\hdots\left( b_{q}\right) _{k}}\frac{x^{k}}{k!}
\end{align}
We obtain: 
\begin{align}
\int_{-1}^{1}dx\left( 1-x^{2}\right) ^{-1}\left[ \Theta\left( x\right)\right]
^{2} & =\frac{\pi}{2^{4k-3}\left( n!\right) ^{2}}\frac{\left(2n+2k+1\right) 
}{\left[ \Gamma\left( k+1\right) \right] ^{2}}\frac { \Gamma\left(
n+2k+1\right) \Gamma\left( 2k-3\right) }{\Gamma\left(k-3/2\right)
\Gamma\left( k-1/2\right) }  \notag \\
& \times_{5}F_{4}\left( 
\begin{array}{c}
-n,k+1/2,n+2k+1,k-1/2,k;1 \\ 
k+1,2k+1,k-1/2,k+1/2%
\end{array}%
\right), \\
\int_{-1}^{1}dx\left( 1-x^{2}\right) ^{-2}\left[ \Theta\left( x\right) %
\right] ^{2} & =\frac{\pi}{2^{4k}\left( n!\right) ^{2}}\frac{\left(
2n+2k+1\right) }{\left[ \Gamma\left( k+1\right) \right] ^{2}}\frac {
\Gamma\left( n+2k+1\right) \Gamma\left( 2k-1\right) }{\Gamma\left(k-1/2%
\right) \Gamma\left( k+1/2\right) }  \notag \\
& \times_{5}F_{4}\left(%
\begin{array}{c}
-n,k+\frac{1}{2},n+2k+1,k-3/2,k-1;1 \\ 
k+1,2k+1,k-3/2,k-1/2%
\end{array}%
\right).
\end{align}

Finally, the diagonal matrix elements up to first order in $\beta $ take the
form: 
\begin{align}
&\frac{\beta }{4\mu }\left\langle Nnm\right\vert p^{4}\left\vert
Nnm\right\rangle =\beta \frac{a^{4}}{\left( n^{\prime }\right) ^{8}}\left\{ 
\frac{\hbar ^{2}}{8\mu }\left( \frac{\hbar ^{2}}{2}+\frac{n^{\prime } }{2l+1}%
\right) \right. +  \notag \\
& +\frac{\eta q\sigma ^{2}\hbar ^{4}}{\mu }\left( \frac{1}{2n^{\prime
}\left( 2l+1\right) }+\frac{4n^{\prime }\Gamma \left( 2l\right) }{%
\Gamma\left( 2l+3\right) }\right) \frac{\pi }{2^{4k-2}\left( n!\right) ^{2}}%
\frac{ \left( 2n+2k+1\right) }{\left[ \Gamma \left( k+1\right) \right] ^{2}}
\notag \\
& \times \frac{\Gamma \left( n+2k+1\right) \Gamma \left( 2k-3\right) }{%
\Gamma \left( k-3/2\right) \Gamma \left( k-1/2\right) }_{5}F_{4}\left( 
\begin{array}{c}
-n,k+1/2,n+2k+1,k-1/2,k;1 \\ 
k+1,2k+1,k-1/2,k+1/2%
\end{array}%
\right)  \notag \\
&+\frac{\eta ^{2}q^{2}\sigma ^{4}\hbar ^{4}}{\mu }\frac{\left[
3\left(n^{\prime }\right) ^{2}-l\left( l+1\right) \right] \Gamma \left(
2l-1\right)}{\Gamma \left( 2l+4\right) n^{\prime }}\frac{\pi }{%
2^{4k+2}\left( n!\right) ^{2}}\frac{\left( 2n+2k+1\right) }{\left[ \Gamma
\left( k+1\right) \right]^{2}}  \notag \\
& \times \frac{\Gamma \left( n+2k+1\right) \Gamma \left( 2k-1\right) }{%
\Gamma \left( k-1/2\right) \Gamma \left( k+1/2\right) }_{5}F_{4}\left(%
\begin{array}{c}
-n,k+\frac{1}{2},n+2k+1,k-3/2,k-1;1 \\ 
k+1,2k+1,k-3/2,k-1/2%
\end{array}%
\right)
\end{align}
This last expression depends on $l(l=n+k)$, which lifts the degeneracy.

\section{Generalized Recurrence Relations}

For the diagonal matrix elements, the radial part verifies the following
recurrence relations given in Ref. \cite{Chang0}(restoring $\mu$, $e$ and $%
\hbar$): 
\begin{equation}
\frac{\hbar ^{4}a^{2}}{4\mu ^{2}e^{4}}\left( s+1\right) \left\langle
r^{s}\right\rangle =\frac{\hbar n^{\prime }a}{2\mu e^{2}}\left\langle
r^{s-1}\right\rangle -\frac{s\left[ \left( 2l+1\right) ^{2}-s^{2}\right] }{4}
\left\langle r^{s-2}\right\rangle\,,  \label{A}
\end{equation}
where the first average elements of $r^{s}$ were evaluated in Ref. \cite%
{Chang0}. Next, we evaluate the recurrence formula for the angular part. For
this we denote: 
\begin{align}
&\left\langle \sin^{2t}\theta\right\rangle _{n,k} =\langle
Nnm\mid\sin^{2t}\theta\mid Nnm\rangle=\int_{-1}^{1}dx\left( 1-x^{2}\right)
^{t}\left[\Theta\left( x\right) \right] ^{2}  \notag \\
& =\frac{\left( 2n+2k+1\right) }{2^{2k+1}\Gamma\left( n+2k+1\right) }\left[%
\frac{\Gamma\left( 2k+1\right) }{\Gamma\left( k+1\right) }\right]%
^{2}\int_{-1}^{1}dx\left( 1-x^{2}\right) ^{k+t}\left[ C_{n}^{\left(
k+1/2\right) }\left( x\right) \right] ^{2}\,.
\end{align}
We can write:%
\begin{align}
&\int_{-1}^{1}dx\left( 1-x^{2}\right) ^{k+t+1}\left[ C_{n}^{\left(k+1/2%
\right) }\left( x\right) \right] ^{2}  \notag \\
&\qquad\qquad=\int_{-1}^{1}dx\left(1-x^{2}\right) ^{k+t}\left( \left[
C_{n}^{\left( k+1/2\right) }\left(x\right) \right] ^{2}-\left[
xC_{n}^{\left( k+1/2\right) }\left( x\right) \right] ^{2}\right).
\end{align}
Using the following recurrence rule of the Gegenbauer polynomials \cite%
{Abramowitz}: 
\begin{equation}
2\alpha\left( 1-x^{2}\right) C_{n-1}^{\left( \alpha+1\right) }\left(x\right)
=\left( 2\alpha+n+1\right) C_{n-1}^{\left( \alpha\right) }\left(x\right)
-nxC_{n}^{\left( \alpha\right) }\left( x\right),
\end{equation}
we straightforwardly obtain: 
\begin{align}
\left\langle \sin^{2\left( t+1\right) }\theta\right\rangle _{n,k} &
=\left\langle \sin^{2t}\theta\right\rangle _{n,k}-\frac{1}{4}\left(
n+2k+1\right) \left\langle \sin^{2\left( t+1\right) }\theta\right\rangle
_{n-1,k+1}  \notag \\
& -\frac{\left( n+2k+2\right) \left( n+2k+1\right) }{n\left( 2n+2k+1\right) }%
\left\langle \sin^{2t}\theta\right\rangle _{n-1,k+1}  \notag \\
& -\frac{\left( n+2k+2\right) }{\left( 2n+2k+1\right) \left( 2n+2k-1\right) }%
\left\langle \sin^{2t}\theta\right\rangle _{n-1,k} \,.  \label{B}
\end{align}

From equations \eqref{A} and \eqref{B} we can derive the general formula of
the averages values of $r^{p}\sin ^{2s}\theta $. The recurrence formula %
\eqref{B} requires the two initial values $\left\langle \sin ^{2t}\theta
\right\rangle _{0,k}$ and $\left\langle \sin ^{2t}\theta \right\rangle _{1,k}
$. Then, taking the special cases (\ref{bb2}) and using of the following
integral \cite{Gradshteyn}: 
\begin{equation}
\int_{0}^{\pi }d\theta \left( z+\sqrt{z^{2}-1}\cos \theta \right) ^{\mu
}\sin ^{2\nu -1}\theta =\frac{2^{2\nu -1}\Gamma \left( \mu +1\right) \left[
\Gamma \left( \nu \right) \right] ^{2}}{\Gamma \left( 2\nu +\mu \right) }%
C_{\mu }^{\left( \nu \right) }\left( z\right) \,,
\end{equation}%
with $\mathrm{Re}(\nu )>0$, we obtain the first matrix elements: 
\begin{align}
\left\langle \sin ^{2t}\theta \right\rangle _{0,k}& =\frac{2^{2t}\Gamma
\left( 2k+2\right) }{\Gamma \left[ 2\left( k+t+1\right) \right] }\left[ 
\frac{\Gamma \left( k+t+1\right) }{\Gamma \left( k+1\right) }\right] ^{2}\,,
\\
\left\langle \sin ^{2t}\theta \right\rangle _{1,k}& =\frac{2^{2t}\left(
2k+3\right) }{2k+2t+3}\frac{\left[ \Gamma \left( 2k+1\right) \right] ^{4}}{%
\Gamma \left( 2k+2\right) \Gamma \left( 2k+2t+1\right) }\left[ \frac{\Gamma
\left( k+t+1\right) }{\Gamma \left( k+1\right) }\right] ^{2}\,.
\end{align}

\section{Conclusion}

In this paper we studied the Schr\"{o}dinger equation for the Hartmann
potential with deformed Heisenberg algebra. Using perturbation theory at the
first order in the parameter of deformation $\beta $, we obtained the
general form of the hamiltonian matrix elements and, as an example, we
showed that the degeneracy of the two states $\left\vert 010\right\rangle $
and $\left\vert 100\right\rangle $ is completely lifted. For the diagonal
matrix elements, we derived an explicit analytical expression which depends
on $l$. In this case, some degenerate states split into sub-levels, and new
transitions appear. In addition to the recurrence formula for the radial
average values given in \cite{Chang0}, we derived the one for the angular
part which leads to the general formula of the average values of $r^{p}\sin
^{2s}\theta $ for the non-relativistic Hartmann potential. These results are
useful in the calculations of the bound-state transitions and, on the
experimental side, the energy levels can be measured and an upper bound on
the minimal length $\left( \Delta X\right) _{\min }$ can be obtained.

\end{document}